\documentclass{elsart}

\usepackage{amssymb}

\begin{document}

\begin{frontmatter}



\title{Sewing Algorithm}


\author{T. E. Booth}
 \address{Applied Physics Division, Los Alamos National Laboratory, Los Alamos, NM 87545}

\author{J. E. Gubernatis}
\address{Theoretical Division, Los Alamos National Laboratory, Los Alamos, NM 87545}

\begin{abstract}
We present a procedure that in many cases enables the Monte Carlo sampling of states of a large system from the sampling of states of a smaller system. We illustrate this procedure, which we call the sewing algorithm, for sampling states from the transfer matrix of the two-dimensional Ising model.

\end{abstract}

\end{frontmatter}

\section{Introduction}
\label{sec:intro}
From its onset, the Monte Carlo method was recognized as an effective approach for estimating the solutions to linear systems of equations, inverting a matrix, and finding eigenvalues of a matrix when these linear algebra problems involved a very large matrix \cite{hammersley}.  The core operation in estimating the solutions of such problems is performing a sequence of matrix-vector multiplications, 
\begin{equation}
\sum_{j,k.\dots,n}A_{ij}A_{jk}\cdots A_{mn}x_n.
\label{eq:product}
\end{equation}
Instead of executing it completely, the Monte Carlo approach samples terms in the summations, for example, $A_{i_1i_2}A_{i_2i_3}\cdots A_{i_{n-1}i_n}x_{i_n}$, by generating a Markov chain  $(i_1,i_2,\dots,i_n)$ defined by a matrix $T_{ij}$  for transition from state $j$ to $i$ and relating the random walks generated by $T_{ij}$ to the sequence matrix-vector multiplications  by having the walker accumulate a ``weight''  $w_{i_1i_2}w_{i_2i_3}\cdots w_{i_{n-1}i_n}$ as it moves along the chain. From the samples, expectation values of the result are easily constructed. There is considerable leeway in choosing the $T_{ij}$. The minimum conditions are
\begin{equation}
A_{ij}=T_{ij}w_{ij}
\label{eq:std}
\end{equation}
where $T_{ij}$ is greater than zero if $A_{ij}$ is nonzero, equals zero if $A_{ij}$ does, and satisfies $\sum_i T_{ij}=1$ for all $j$. The core Monte Carlo sampling is for a given $j$ selecting an $i$ with the conditional probability $T_{ij}$. 

Sampling the conditional probability is straightforward until the matrix becomes too large to store or too expensive to continually regenerate. It is these situations that this paper addresses by presenting a new strategy, the sewing algorithm, which should be applicable to a variety of problems.  The sewing algorithm accomplishes the sampling  of states $i$ for a large system from samples of states from transition matrices of small systems. It was developed in the context of a power method algorithm to obtain multiple eigenvalues of very large matrices \cite{booth1,booth2, gubernatis,booth08}. In this eigenvalue work the benchmark application was obtaining the two largest eigenvalues of the transfer matrix of the two-dimensional Ising model \cite{thompson}. With the Sewing Algorithm we easily obtained these two eigenvalues for a $60 \times 60$ lattice on a single processor. Past work required parallel computing. $60 \times 60$ is where we decided to stop and likely not the limit of the algorithm. Reference \cite{booth08} contains details about the application and references to the eigenvalue algorithms.

We will present the sewing algorithm in the context of the transfer matrix for the two-dimensional Ising model. While some of the details will be relevant only to this and related problems, the basic strategy is more general. In the next section, Section 2, we will define the transfer matrix \cite{montroll}. In Section 3, we will discuss the basics for implementing the Monte Carlo to sample a sequence of matrix-vector multiplications when the system is still small enough so that there is enough computer memory to store $T_{ij}$. Then, in Section 4, we will discuss how to sample $T_{ij}$, when memory is inadequate for the system size of interest but is adequate for smaller system sizes. Finally, in the last section, we will make some comments about the generality of the method.

\section{Transfer Matrix of Two-Dimensional Ising Model}
\label{sec:ising}
We will consider an $m\times m$ Ising model defined with periodic boundary conditions in one direction and open boundary conditions in the other.  The model's energy is \cite{thompson}
\begin{equation}
E\left\{ \mu  \right\} =  - J\sum\limits_{i = 1}^{m-1} {\sum\limits_{j = 1}^m {\mu _{i,j} \mu _{i+1,j } } }  -
J\sum\limits_{i = 1}^m {\sum\limits_{j = 1}^m {\mu _{i,j} \mu _{i,j + 1} } }
\label{eq:energy}
\end{equation}
Here, $(i,j)$ are the coordinates of a lattice site. The Ising spin variable $\mu_{i,j}$ on each site has the value of $ \pm 1$, the exchange constant $J>0$,  and $\mu _{i,m + 1}  = \mu _{i,1}$. The symbol 
\begin{equation}
\sigma _j  = \left( {\mu _{1,j} ,\mu _{2,j} , \ldots ,\mu _{m,j} } \right)
\label{eq:state}
\end{equation}
denotes a column configuration of Ising spins and there are $2^m$ possible configurations for each column. Typically, a configuration is mapped onto an integer in the range $0$ to $2^m-1$ with a $+1$ Ising spin mapped to a 1 bit in the integer and a $-1$ spin to a 0 bit.  
The transfer matrix $A(\sigma,\sigma')$  follows from a re-expression of the partition function \cite{thompson}
\begin{eqnarray}
 Z\left( {m,m} \right) & = &\sum\limits_{\left\{ \mu  \right\}} {\exp \left[ { - \nu E\left( {\left\{ \mu  \right\}} \right)} \right]}  \nonumber \\
  & = &\sum\limits_{\sigma _1 , \ldots ,\sigma _m } {\exp \left[ {\nu \left( {\sum\limits_{j = 1}^m
  {\left\{ {S_1 \left( {\sigma _j } \right) + S_2 \left( {\sigma _j ,\sigma _{j + 1} } \right)} \right\}} } \right)} \right]}  \nonumber \\
  & = & \sum\limits_{\sigma _1 , \ldots ,\sigma _m } {A(\sigma _1 ,\sigma _2 )} A(\sigma _2 ,\sigma _3 )
  \cdots A(\sigma _{m - 1} ,\sigma _m )A(\sigma _m ,\sigma _1 ) \nonumber \\
  & = &\sum\limits_{\sigma _1 } {A^m (\sigma _1 ,\sigma _1 )}  
\end{eqnarray}
where
\begin{equation}
S_1 \left( {\sigma _j } \right) =  \sum\limits_{i = 1}^{m - 1} {\mu _{i,j} \mu _{i + 1,j} }
\label{eq:s1}
\end{equation}
is the interaction energy of the $j^{th}$ column and
\begin{equation}
S_2 \left( {\sigma _j ,\sigma _{j + 1} } \right) =  \sum\limits_{i = 1}^m {\mu _{i,j} \mu _{i,j + 1} }
\label{eq:s2}
\end{equation}
is the interaction energy between the $j^{th}$ and $(j+1)^{th}$ columns, $\nu=J/k_B T$, $k_B$ is Boltzmann's constant, and $T$ is the temperature.  $A(\sigma,\sigma')$ is a $2^m \times 2^m$ matrix whose elements are
\begin{equation}
A\left( \sigma ,\sigma'\right) = \exp \left(\nu \sum\limits_{k = 1}^{m -1}
\mu _k \mu _{k + 1} \right)  \exp \left(\nu \sum\limits_{k = 1}^m
\mu _k \mu _k^{'} \right)
\label{eq:transfer_matrix}
\end{equation}
Because of the one open boundary, the matrix is asymmetric. We note that all the elements of $A(\sigma,\sigma')$ are greater than zero.  The mapping of the $\sigma$ to integers maps $A\left( \sigma ,\sigma'\right) $ to $A_{ij}$.

\section{Core Monte Carlo}
\label{sec:mc}
We assume that the elements of some $M\times M$ matrix $A$ are easily generated on-the-fly. For simplicity, but without loss of generality, we will also assume that all its elements are positive. Next, we imagine we have $N$ random walkers distributed over the $M$ states defining $A$, and we will interpret $A_{ij}$ as the weight of particles arriving in state $i $ per unit weight of a walker in state $j$ and will regard the action of $A$ on  $x$ as causing a walker to jump from some $j$ to some $i$, carrying its current weight $x_j$, modified by $A_{ij}$, to state $i$.  Repeated walks by a walker and the sum over all walkers estimates (\ref{eq:product}).

The jumps are executed probabilistically. To do this, instead of (\ref{eq:std}), we define  the total weight leaving state $j$ as
\begin{equation}
W_j=\sum_i A_{ij}
 \label{eq:weight_multiplier}
\end{equation}
and the transition probability from $j$ to $i$ as
\begin{equation}
T_{ij}=A_{ij}/W_j
 \label{eq:transition_probability}
\end{equation}
The number $W_j$ is called the state weight multiplier. 

If $M$ is  sufficiently small, then a Monte Carlo procedure is easily constructed. We sample a state $i$ from $T_{ij}$ and multiply the transferred weight by the ratio of the true probability (1.0) to the sampled probability ($T_{ij}$); that is, if state $i$ is sampled, the weight arriving in state $i$ from state $j$ is multiplied by
 \begin{equation}
A_{ij} \frac{1.0}{T_{ij}}=A_{ij} \frac{1.0}{A_{ij}/W_j}=W_j
\end{equation}
The standard way \cite{kalos} to sample $i$ from $T$ is to first construct the cumulative distribution,
\begin{equation}
C_{ij}=\sum_{k=1}^{i}T_{kj}
\end{equation}
and then draw a random number $\xi$ from the unit distribution. The state $i$ equals the value of $k$ that satisfies
\begin{equation}
C_{k-1,j}< \xi \le C_{k,j}
\end{equation}
where $C_{0j}=0$.

\section{Sewing Algorithm}
\label{sec:sewing}
For $M$ small enough so $C_{ij}$ can be stored in memory, sampling from the cumulative probability $C_{ij}$ works well. If the number of states gets too large, then $C_{ij}$ cannot be sampled directly.  In this case, instead of, for example, just randomly picking from any state $j$ any state $i$ with probability $1/M$, we developed a new procedure which we call the sewing algorithm.  In it, we assume, for example, that we can write any state $i$ as a direct product of the states in a smaller basis. If it is the direct product of two states, that is,
\begin{equation}
\left| i \right\rangle  = \left| {i_2 } \right\rangle \left| {i_1 } \right\rangle 
\end{equation}
then instead of transferring weight
\begin{equation}
W_j=\sum_k A_{kj}
\end{equation}
from state $j$ to state $i$ with probability
\begin{equation}
T_{ij}=A_{ij}/\sum_k A_{kj}=A_{ij}/W_j,
\end{equation}
we will use the $a_{ij}$ that would apply to the smaller set of states and then make an appropriate weight correction.

For each smaller set of states, we have
\begin{eqnarray}
t_{ij} &=& a_{ij}/w_j,\\
\label{eq:trans_mtrx}
w_j &=& \sum_k a_{kj}
\label{eq:wgt_mult}
\end{eqnarray}
and for $\left| j \right\rangle  = \left| {j_2 } \right\rangle \left| {j_1 } \right\rangle $ we sample $|i_1\rangle$ and $|i_2\rangle$ from
\begin{equation}
  t_{i_2j_2}   t_{i_1j_1}
\label{eq:trans_prob}
\end{equation}
Now, we define $X_{ij}$ to be the weight correction necessary to preserve the expected weight transfer from state $j$ to state $i$. It satisfies
\begin{equation}
A_{ij}=  X_{ij}  t_{i_2j_2}   t_{i_1j_1}
\label{eq:transfer_mtrx}
\end{equation}
Thus
\begin{equation}
A_{ij}=  X_{ij}  \frac{a_{i_2j_2}}{w_{j_2} }
                         \frac{a_{i_1j_1}}{w_{j_1} }
\label{eq:mtrx}
\end{equation}
that is
\begin{equation}
X_{ij} = w_{j_1} w_{j_2} \frac{A_{ij}} {a_{i_1j_1} a_{i_2j_2} }
\label{eq:weight_correction}
\end{equation}

This sewing method generalizes easily. For $k$ sets of states, (\ref{eq:mtrx})  and (\ref{eq:weight_correction})  become
\begin{equation}
A_{ij}=  X_{ij}  \prod_{n=1}^k t_{i_nj_n}
\end{equation}
and
\begin{equation}
X_{ij}=A_{ij} \prod_{n=1}^k\frac{w_{j_n}}{a_{i_nj_n}}
\end{equation}

\subsection{Illustration for the Transfer Matrix}
For the Ising model
\begin{equation}
A_{ij}=\exp[\nu (S_1(i)+S_2(i,j))]
\end{equation}
where $S_1$ and $S_2$ were defined in (\ref{eq:s1}) and (\ref{eq:s2}). To compute the weight correction factor $X_{ij}$, we need to study how the sums $S_1(i)$ and $S_2(i,j)$ differ between evaluating them with the bits taken together and taken separately.
 
We will establish this difference in the context of the following assumed binary representation of two states
\begin{eqnarray}
| i \rangle &=&|10101000110011011110\rangle
            = |1010100011\rangle |0011011110\rangle    \label{eq:i} \\
| j \rangle&=&|10101100111011011110\rangle
            = |1010110011\rangle |1011011110\rangle   \label{eq:j}
 \end{eqnarray}
The sum $S_1(i)$, computed on $i$ alone, adds 1 every time adjacent bits match and subtracts 1 every time they mismatch. Because of the periodic boundary condition, the two ends are adjacent. For (\ref{eq:i}), there are 12 matches and 8 mismatches so $S_1(i)=-4$. The sum $S_2(i,j)$ counts matches and mismatches between the bits of $i$ and $j$. For (\ref{eq:i}) and (\ref{eq:j}), there are 18 matches and 2 mismatches so $S_2(i,j)=16$. The transfer matrix element from $j$ to $i$ is then 
\begin{equation}
  A_{ij}=\exp[\nu (S_1(i)+S_2(i,j))]=\exp(12\nu)
\end{equation}

In sewing the bits together, our intent was to proceed recursively so we asumed periodic boundary conditions to hold for the various bit segments; that is, $a_{ij}$ was the transfer matrix of a smaller-sized Ising model. First, we will consider just the lower 10 bits of $i$.
\begin{equation}
| i_1 \rangle = |0011011110\rangle
\end{equation}
There are 6 matches and 4 mismatches so its first sum $s_1(i_1)=2$. For the upper 10 bits of $i$,  
\begin{equation}
| i_2 \rangle = |1010100011\rangle
\end{equation}
there are 4 matches and 6 mismatches so its first first sum $s_1(i_2)=-2$

We note that the terms making up $S_1(i)$ are {\em almost} the same as the terms making up $s_1(i_1)$ and $s_1(i_2)$. The differences occur at the ends of the sets of bits. In the present example the difference between taking bit segments together rather than separately are two mismatches occurring  at the right ends of each segment. Thus $S_1(i)=(s_1(i_1) -2) + (s_1(i_2) -2) =-4$. We now write $ S_1(i)=s_1(i_1)+s_1(i_2)+D(i)$ where $D(i)$ is defined to be the ``energy'' difference between calculating the sets of bits together and calculating the sets of bits separately.

For $S_2(i,j)$ the number of matches and mismatches between $i$ and $j$ is just the sum of the matches and mismatches on the sewn pieces, that is, $S_2(i,j)=s_2(i_1,j_1)+s_2(i_2,j_2)$.
 
We can now write 
\begin{equation}
X_{ij} = w_{j_{2}} w_{j_{1}} \frac{\exp[\nu (S_1(i)+S_2(i,j))]}
{\exp[\nu (s_1(i_{2})+s_2(i_{2},j_{2}))]
  \exp[\nu (s_1(i_{1})+s_2(i_{1},j_{1}))] }
\label{eq:cij}
\end{equation}
We showed that 
\begin{equation}
S_2(i,j)=s_2(i_{2},j_{2})+s_2(i_{1},j_{1})
\label{eq:xs1}
\end{equation}
and 
\begin{equation}
S_1(i)=s_1(i_{2})+s_1(i_{1})+D(i)
\label{eq:xs2}
\end{equation}
Substituting (\ref{eq:xs1})  and (\ref{eq:xs2}) into  (\ref{eq:cij})  yields
\begin{equation}
X_{ij} = w_{j_{1}} w_{j_{2}} \exp[\nu D(i)]
\end{equation}

The weight correction for sewing $k$ systems together becomes
\begin{equation}
X_{ij} =\exp(\nu D_i)\prod_{n=1}^k w_{j_n}
\end{equation}
where $\nu D_i$ is the energy difference per $k_BT$ between calculating with the bits together and the bits separately.

\section{Concluding Remarks}
\label{sec:concluding}
The sewing algorithm is very simple and precise in concept, and its application should extend beyond the specific case illustrated. Our presentation had two classes of assumptions, one for the matrices and the other for the states.

We assumed all our matrix elements were positive. If some are zero, then the corresponding $T_{ij}$ must be zero.  If some are negative, the basic strategy prevails when absolute values are taken in the appropriate places. Mixed signed matrix elements necessitates mixed signs of the weights of the walkers, and additional Monte Carlo procedures might be necessary to promote proper cancellations of the signs. This need was the case for our work in computing multiple eigenvalues of large matrix via the Monte Carlo method \cite{booth08}.

We also assumed the state of the larger system was expressible as a direct product of the states of a smaller systems. Clearly, this specific requirement was possible and convenient for the Ising problem and is likely so for other problems involving interacting  spins, electrons, and hard-core bosons. More generally, the larger system simply needs to be domain-like decomposable with interfacial corrections easily computed. The details for doing this will depend of the specific problem. Many, but not all, problems will be efficiently amendable to the sewing algorithm.



\end{document}